\documentclass[prd,a4paper,superscriptaddress,nofootinbib,10pt]{revtex4} 
\usepackage{amsfonts}
\usepackage{amssymb}
\usepackage{amsmath}
\usepackage{marvosym}
\usepackage{txfonts}
\usepackage[T1]{fontenc}
\usepackage{mathtools}
\usepackage{lscape}
\usepackage{mathrsfs}
\usepackage{graphicx}
\usepackage{bm}
\usepackage{color}
\usepackage{slashed}
\usepackage{hyperref}
\usepackage{tikz}
\usepackage{pgfplots}
\usepackage{array}
\usepackage{textcomp}
\usepackage{epsfig}
\usepackage{slashed}
\usepackage{comment}
\usepackage{xcolor}
\usepackage{enumerate}
\pgfplotsset{width=10cm}

\begin{document}

\title{Entropy of black holes, charged probes and noncommutative generalization}
\author{Axel Hrelja}
\email{ahrelja@irb.hr}
\affiliation{Rudjer Bo\v{s}kovi\'{c} Institute, Bijen\v{c}ka c.54, HR-10002 Zagreb, Croatia}

\author{Tajron Juri\'c}
\email{tjuric@irb.hr}
\affiliation{Rudjer Bo\v{s}kovi\'{c} Institute, Bijen\v{c}ka c.54, HR-10002 Zagreb, Croatia}

\author{Filip Po\v{z}ar}
\email{fpozar@irb.hr}
\affiliation{Rudjer Bo\v{s}kovi\'{c} Institute, Bijen\v{c}ka c.54, HR-10002 Zagreb, Croatia}

\date{\today}

\begin{abstract} 
The brick wall model is a semi-classical approach to understanding the microscopic origin of black hole entropy. We outline the formalism for the brick wall model in arbitrary number of dimensions and generalize it to include both charged spacetimes and charged probes in order to systematically show how to calculate the entropy for any black hole, in higher orders of the WKB approximation. We calculate the entropy for the Reissner-Nordstr\"{o}m and charged BTZ black holes, and by looking at the chargeless limits we recover the entropy of the Schwarzschild and neutral BTZ black holes. We also study noncommutative corrections to the black hole entropy by using a Drinfeld twist to deform spacetime symmetries. Using the noncommutative action for a charged scalar field, we derive the noncommutative Klein-Gordon equation and the radial equation in arbitrary dimension on which we generalize the brick wall method. We study case by case the noncommutative Reissner-Nordstr\"{o}m and noncommutative charged BTZ black holes, the entropy of which we calculate using the generalized brick wall method. Corrections due to higher order in the WKB approximation arise, and they are given by the logarithms of the black hole area. The corrections in the lowest order of WKB due to noncommutativity are also given by the logarithms of the black hole area.
\end{abstract}

\maketitle

\section{Introduction}

Black hole entropy was first introduced by Bekenstein \cite{Bekenstein:1972tm} to preserve the universal applicability of the second law of thermodynamics. After the formalization of the thermodynamic properties of black holes as four laws of black hole mechanics \cite{Bardeen:1973gs}, Hawking \cite{Hawking:1975vcx} showed that a black hole emits thermal radiation at a fixed temperature, which lead to fixing the constant of proportionality between the entropy and the black hole area, known as the Bekenstein-Hawking law. Later the origin of the entropy was understood through statistical mechanics of in-falling particles \cite{Zurek:1985gd} or scalar fields by using a brick wall to regulate the ultraviolet divergence \cite{tHooft:1984kcu, tHooft:1996rdg}. There are various approaches to quantum gravity where the entropy of black holes corresponds with the Bekenstein-Hawking formula, such as string theory \cite{Strominger:1996sh, Sen:2012kpz, Sen:2012dw}, quantum geometry \cite{Ashtekar:1997yu}, conformal field theory \cite{Strominger:1997eq, Carlip:1999cy, Carlip:2000nv, Gupta:2001bg}, AdS/CFT duality \cite{Mukherji:2002de}, higher loop corrections \cite{Fursaev:1994ea, Demers:1995dq, Solodukhin:2019xwx, Solodukhin:2011gn}, nonlocal effective field theories \cite{Xiao:2021zly, Calmet:2021lny}, entanglement entropy \cite{Bombelli:1986rw, Srednicki:1993im}, noncommutative geometry \cite{Doplicher:1994zv, Doplicher:1994tu, Szabo:2001kg, Chaichian:2004za, Aschieri:2005yw, Aschieri:2005zs, Ohl:2008tw, Ohl:2009pv, Moffat:2000gr, Chamseddine:2000zu, Nishino:2001gt, Balachandran:2006qg, Harikumar:2006xf, Bastos:2007bg, Bastos:2009ae}, etc.\\

In this paper, motivated by the extension of the brick wall in arbitrary dimensions \cite{Sarkar:2007uz}, we generalize the method to charged spacetimes and charged probes. We calculate the entropy of the Reissner-Nordstr\"{o}m and charged BTZ (QBTZ) black holes up to the second order in the WKB approximation. Looking at the limits of chargeless spacetimes and chargeless probes, we arrive at entropies for the Schwarzschild and BTZ black holes. \\

After generalizing the brick wall method from \cite{Sarkar:2007uz} to arbitrary dimensional charged black holes and charged scalar probes, we further generalize the method to more general Schrodinger-like radial equations arising from actions defined on noncommutative spaces. We found the general expressions, in all spacetime dimensions, for all the terms in the modified radial equation of motion for the so called  angular twist \cite{Ciric:2017rnf} and present the integral for the entropy. Setting the charges of black hole and scalar probe to zero, along with setting the noncommutativity parameter to zero reproduces the results of \cite{Sarkar:2007uz}.\\

Noncommutative geometry in field theory 
can be implemented by the means of (Drinfeld) twisting symmetry of the theory in the Hopf Algebra approach, e.g., by deforming diffeomorphism \cite{Konjik:2020fle, Aschieri:2009ky, Herceg:2023zlk}, Poincar\'e \cite{Gupta:2011uz, Govindarajan:2009wt, Dimitrijevic:2011jg, Lukierski:1993hk,Gupta:2013ata} and gauge symmetry \cite{Ciric:2017rnf, DimitrijevicCiric:2019hqq, Konjik:2020fle, Dimitrijevic:2011jg,Seiberg:1999vs,Aschieri:2011ng,  Gupta:2022oel, Juric:2022bnm}. Noncommutativity of spacetime was also modeled in a variety of other ways which are based on other methods \cite{Chaichian:2000ia, Chaichian:2001nw, Lukierski:1991pn, Lukierski:1992dt, Lukierski:1993wxa, Dolan:2006hv, Schupp:2009pt, Jurco:2001rq, Mureika:2011py, Anacleto:2020zfh, Banerjee:2008gc}, but in this paper we only consider noncommutativity introduced via the Drinfeld twisted symmetry. In the commutative setting, the pointwise algebra $\left(C^\infty(\mathcal{M}),\cdot\right)$ is covariant under the undeformed symmetry Hopf algebra. On the other hand, in the deformed/twisted case, one also needs to deform the pointwise algebra into a star-product algebra $\left(C^\infty(\mathcal{M}),\star\right)$ in order to have a well defined action of the deformed symmetry Hopf algebra on spacetime functions. Having said that, the noncommutative action needs to be defined with star products in order to obey deformed symmetry. In this paper we will analyze systems of charged black holes and charged probes, so we are going to consider noncommutative $U(1)_\star$ gauge theory. For the $U(1)_\star$ scalar theory, the action is given as
\begin{equation}
	S [\hat{\Phi}] = \int \sqrt{-g}\left[g^{\mu\nu}\left( \hat{D}_\mu \hat{\Phi}\right)^+ \star \left( \hat{D}_\nu \hat{\Phi}\right) - \frac{m^2}{\hbar^2} \hat{\Phi}^+\star \hat{\Phi} \right]\;,
\end{equation}\\
where the $\star$-product  is  the Moyal product
\begin{equation}
	(f \star g)(x)=\lim_{y\to x}\left( e^{-\frac{i}{2}\Theta^{\mu\nu}\partial^x_\mu \partial^y_\nu} f(x)g(y)\right),
\end{equation}
and $\hat{D}_\mu$ is the gauge-covariant derivative
\begin{equation}
	\hat{D}_\mu \hat{\Phi} = \partial_\mu\hat{\Phi} + i\frac{q}{\hbar}\hat{A}_\mu\star\ \hat{\Phi}\;.
\end{equation}
Finally, when dealing with noncommutative gauge symmetry, in order to obtain the equation of motion on which to apply the brick wall procedure on, it will be required to expand the fields with the Seiberg-Witten map alongside expanding the star product and keeping the lowest order corrections in $\Theta$. \\

This paper is organized as follows. In section II we review the brick wall method in arbitrary dimensions for a spherically symmetric spacetime \cite{tHooft:1984kcu, Sarkar:2007uz} and then generalize the calculation to charged black holes and charged probes. 
In Section III we calculate the entropy for the most simple black holes up to second order. Specifically, in Section IIIA we calculate the entropy of the Reissner-Nordstr\"{o}m (RN) black hole with a charged probe up to the second WKB order. In Section IIIB we do so for the charged BTZ (QBTZ) black hole. Section IIIC deals with the limiting cases of chargeless black holes and chargeless probes for the entropies calculated in IIIA and IIIB, while IIID does so only for chargeless probes. In Section IV we deal with the noncommutative (NC) corrections to the entropy of the RN black hole and the QBTZ black hole. Our analysis in the Sections III and IV demonstrate that it is possible to obtain logarithmic corrections to the black hole entropy even in the lowest WKB order. In Section V we conclude the paper with some final remarks.\\

Throughout the paper we are using the units in which $k_B = c = G = 1$, while keeping $\hbar$ since we will use the WKB approximation and expansions in $\hbar$.

\section{Charged generalization of the brick wall method}

The brick wall model is a semi-classical approach to understanding the microscopic origin of black hole entropy. The black hole geometry is assumed to be a fixed classical background on which matter fields propagate, and the entropy of black holes arises due to the canonical entropy of matter fields outside the black hole event horizon, evaluated at the Hawking temperature.
Considering the number of energy levels that a particle can occupy in the vicinity of a black hole, one finds that the density of states diverges \cite{tHooft:1984kcu}. To remedy this UV divergence, 't Hooft introduced a so-called brick wall near the horizon by assuming that all fields $\Phi$ must vanish within some fixed distance $h$ from the horizon,
\begin{equation}
\Phi(r)=0, \quad r= r_\mathrm{H}+h,
\end{equation}
where $r_\mathrm{H}$ is the position of the horizon of the black hole.

 This method was extended in \cite{Sarkar:2007uz} to a $(D+2)$-dimensional, spherically symmetric black hole spacetime with the metric
\begin{equation}
ds^2=-f(r)dt^2+\frac{dr^2}{g(r)}+r^2 d\Omega_D^2,
\label{metrika}
\end{equation}
where $d\Omega_D^2$ is the metric of a $D-$sphere. In what follows let us recap\footnote{We will recap the derivations form \cite{Sarkar:2007uz} with more detail due to the typos in equations (27, 28, 37) that arose in that paper.} the method from \cite{Sarkar:2007uz}. The surface gravity $\kappa$ is defined as
\begin{gather}
    k^a\nabla_a k^b=\kappa k^b,
\end{gather}
where $k^a$ are Killing vectors defined by the Killing equation
\begin{gather}
    \nabla_{a}k_{b}+\nabla_b k_a=0.
\end{gather}
Contracting the following property of the Killing vectors
\begin{gather}
    k_a\nabla_b k_c + k_b \nabla_c k_a + k_c \nabla_a k_b=0
\end{gather}
from the left with $\nabla^a k^b$, and evaluating it for the time-like Killing vector, we arrive at a formula for the surface gravity
\begin{gather}
    \kappa=\left.\sqrt{-\frac{1}{2}(\nabla^ak^b)(\nabla_a k_b)} \right|_{r=r_\mathrm{H}}=\left.\left[\sqrt{\frac{g(r)}{f(r)}} \left(\frac{f'(r)}{2}\right) \right]\right|_{r=r_\mathrm{H}}.
\end{gather}
When evaluating the contributions to the brick wall entropy at the higher orders in the WKB approximation, we find that it is useful  to expand the metric functions $g(r)$ and $f(r)$ near the horizon up to second order,
\begin{gather}
    f(r)=f'(r_\mathrm{H})(r-r_{\mathrm{H}})+\frac{1}{2}f''(r_\mathrm{H})(r-r_\mathrm{H})^2+\ldots, \\\nonumber
    g(r)=g'(r_\mathrm{H})(r-r_{\mathrm{H}})+\frac{1}{2}g''(r_\mathrm{H})(r-r_\mathrm{H})^2+\ldots,
\end{gather}
where the horizon is defined as the value at which the metric functions vanish 
\begin{gather}
    g(r_\mathrm{H})=f(r_\mathrm{H})=0.
\end{gather}
Another useful quantity to define is the proper (coordinate invariant) radial distance from the event horizon of the black hole to the brick wall which is defined as
\begin{gather}
    h_c=\int_{r_\mathrm{H}}^{r_\mathrm{H} +h} \frac{dr}{\sqrt{g(r)}}.
\end{gather}
Expanding $g(r)$ up to first order we are left with
\begin{gather}
    h_c = \sqrt{\frac{4h}{g'(r_\mathrm{H})}}.
    \label{hc}
\end{gather}
The scalar field $\Phi$ satisfies the Klein-Gordon equation
\begin{equation}
    \left(\Box - \frac{\mu^2}{\hbar^2} \right) \Phi =0,
\end{equation}
where $\mu$ is the mass of the scalar field, and $\Box=\nabla^\mu \nabla_\mu$ represents the d'Alembertian operator which can be rewritten as
\begin{align}
\Box  = \frac{1}{\sqrt{-g}}\partial_\mu(g^{\mu \nu}\sqrt{-g}\partial_\nu) = \frac{1}{\sqrt{-g}}\partial_t(g^{tt}\sqrt{-g}\partial_t) + \nabla^2_{D+1},
\end{align}
where $\nabla_{D+1}^2$ is the $(D+1)$-dimensional Laplace operator in curved spacetime, which in the case of a diagonal metric has the form
\begin{equation}
\nabla_{D+1}^2=\frac{1}{\sqrt{-g}}\partial_i(g^{ij}\sqrt{-g}\partial_j) = \frac{1}{\sqrt{-g}} \sum_{i=1}^{D+1} \partial_i(g^{ii}\sqrt{-g}\partial_i). 
\end{equation}
Since the metric is diagonal, its determinant $g$ has the form
\begin{equation}
g=-\frac{f(r)}{g(r)}r^{2D}H(\theta,\phi,...),
\end{equation}
where $H$ is a function of all angle variables of the $D-$sphere.
The Klein-Gordon equation can now be expressed as
\begin{equation}
\left(-\frac{1}{f(r)}\partial_{tt}+\nabla_{D+1}^2-\frac{\mu^2}{\hbar^2}\right)\Phi=0.
\label{komutativni KG}
\end{equation}
Due to the rotational symmetry of the spacetime (\refeq{metrika}) we can use the ansatz \cite{Frolov:1989rv}
\begin{equation}
\Phi=e^{-iEt/\hbar}\frac{R(r)}{r^{D/2}\sqrt{G(r)}}Y_{\ell m_i}(\theta, \phi_i),
\label{ansatz}
\end{equation}
where $G(r)=\sqrt{f(r) g(r)}$, $i\in \{1,...,(D-1)\}$ and $Y_{\ell m_i}(\theta, \phi_i)$ denote the hyperspherical harmonics. By plugging (\refeq{ansatz}) into (\refeq{komutativni KG}), and using \cite{Frolov:1989rv}
\begin{equation}
    \nabla_{D+1}^2Y_{\ell m_i}(\theta, \phi_i)=-\frac{\ell(\ell+D-1)}{r^2}Y_{\ell m_i}(\theta, \phi_i),
\end{equation}
the radial equation is given by
\begin{equation}
    R''(r)+\left[\frac{V^2(r)}{\hbar^2}-\Delta(r)\right]R(r)=0,
\end{equation}
where $V^2(r)$ and $\Delta(r)$ are equal to
\begin{equation}
V^2(r)=\frac{1}{G(r)^2}\left(E^2-g(r)\bigg[\mu^2+\left(\frac{\ell(\ell+D-1)\hbar^2}{r^2}\right)\bigg]\right),
\label{V2}
\end{equation}
\begin{equation}
    \Delta(r)=\left(\frac{D(D-2)}{4 r^2} + \frac{D}{2 r}\frac{G'(r)}{G(r)} - \frac{(G'(r))^2}{4 G(r)^2} + \frac{G''(r)}{2 G(r)}\right).
    \label{Delta}
\end{equation}
We can notice that the $V^2(r)$ term plays the role of the effective potential \cite{Sarkar:2007uz}. Since it is difficult to find an exact analytical solution for the function $R(r)$, we resort to using the WKB approximation
\begin{gather}
    R(r)=\frac{1}{\sqrt{P(r)}} \exp{\left[\frac{i}{\hbar}\int^r P(r') dr'\right]}\;,
\end{gather}
and we arrive at the following differential equation for $P(r)$
\begin{gather}
    P(r)^2[P(r)^2-V^2(r)]=\hbar^2\left[\frac{3}{4}{P'(r)}^2-\frac{1}{2}P''(r)P(r)-\Delta(r){P(r)}^2\right].
    \label{WKB ODJ}
\end{gather}
Following the analysis of \cite{Sarkar:2007uz}, we expand $P(r)$ in powers of $\hbar^2$
\begin{equation}
P(r)=\sum_{n=0}^{\infty}\hbar^{2n}P_{2n}(r).
\end{equation}
Inserting this series into (\refeq{WKB ODJ}), and grouping the terms by powers of $\hbar^2$, we arrive at the equations for $P_{2n}(r)$, up to $n=2$
\begin{equation}
    \begin{split}
    P_0(r) &=\pm V(r), \\
    P_2(r) &=\left(\frac{3}{8P_0(r)}\right)\left(\frac{P_0'(r)}{P_0(r)}\right)^2-\left(\frac{P_0''(r)}{4P_0(r)^2}\right)-\left(\frac{\Delta(r)}{2P_0(r)}\right), \\
    P_4(r) &=-\left(\frac{5P_2(r)^2}{2V(r)}\right)-\left(\frac{4P_2(r)\Delta(r)+P_2''(r)}{4V^2(r)}\right)+\left(\frac{3P_2'(r)V'(r)-P_2(r)V''(r)}{4V^3(r)}\right)\;,
    \label{P komutativni}
    \end{split}
\end{equation}
where we will only consider the $+$ branch of $P_0(r)$. Notice how higher order terms $P_{2n}(r)$ for $n>1$ can be recursively written as functions of $P_0(r)$. Furthermore, $P(r)$ was written as a series in $\hbar^2$ instead of $\hbar$ since the terms proportional to odd powers of $\hbar$ vanish. This can be easily shown by plugging an expansion that includes odd powers into the equation (\refeq{WKB ODJ}) and grouping the terms in powers of $\hbar$.
As long as $P_{2n}(r)$ is real, we can impose the Bohr-Sommerfeld quantization condition
\begin{equation}
\pi n= \frac{1}{\hbar} \int_{r_{\mathrm{H}}+h}^{L}dr P(r),
\label{Sommerfeld}
\end{equation}
which gives the quantization of energy. The total number of wave solutions with the energy less than or equal to $E$ is given by
\begin{gather}
    N(E)= \sum_{\ell=0}^{\ell_{\text{max}}}\sum_{m=-\ell}^\ell\mathcal{W}(\ell) n \longrightarrow  \frac{\hbar^{2n-1}}{\pi}\sum_{n=0}^\infty\int_{r_{\mathrm{H}}+h}^L dr \int_0^{\ell_{\text{max}}} d\ell (2\ell+D-1) \mathcal{W}(\ell) P_{2n}(r) \equiv \sum_{n=0}^\infty N_{2n}(E),
    \label{broj stanja}
\end{gather}
where $N_{2n}(E)$ is the contribution of the $n$-th WKB mode to the total number of states of the field with energy less than $E$, and $\mathcal{W}(\ell)$ is defined as
\begin{equation}
\mathcal{W}(\ell)=\frac{(\ell+D-2)!}{(D-1)!\ell!}.
\end{equation}
Note that we approximated the sum over the angular quantum numbers $\ell$ as an integral, with a degeneracy factor $\mathcal{W}(\ell)$ which is relevant at spacetime dimensions that differ from $D=2$. This approximation is considered to be valid since the separation between the states are expected to be small $(\ell_{\text{max}}\gg 1)$. $\ell_{\text{max}}$ is given in such a way that the functions $P_{2n}(r)$ are real, or equivalently, that $P_0(r)$ is real. The condition on $\ell_{\text{max}}$ is then
\begin{equation}
\ell_{\text{max}}(\ell_{\text{max}}+D-1)=\frac{r^2}{\hbar^2}\left(\frac{E^2}{g(r)}-\mu^2\right).
\label{lmax}
\end{equation}
We should note that the upper bound of the radial integral (\refeq{Sommerfeld}), $L$, signifies the infrared cutoff, imposed to guarantee the finiteness of the entropy at large distances.
We decompose the entropy and the free energy in the same manner
\begin{equation}
S=\sum_{n=0}^{\infty}S_{2n}, \quad F=\sum_{n=0}^{\infty}F_{2n},
\end{equation}
where
\begin{gather}
	F_{2n}=-\int_0^{\infty}\frac{N_{2n}(E)}{e^{\beta E}-1}dE, \quad S_{2n}=\beta^2 \frac{\partial F_{2n}}{\partial \beta}.
	\label{F2n S2n}
\end{gather}
Here the entropy is evaluated at the Hawking temperature
\begin{equation}
	T_H = \frac{1}{\beta_H} = \kappa \frac{\hbar}{2\pi} = \frac{g'(r_H)\hbar}{4\pi}\;.
	\label{Hawkingova temperatura}
\end{equation}
We can now calculate the entropy of a black hole up to an arbitrary order of the WKB approximation by only knowing its dimension and its metric. Furthermore, we are interested in the main contribution to the total entropy coming from the horizon. The radial integral in the entropy, after switching to the near-horzion coordinate, $x=r-r_\mathrm{H}$ can be split into two parts 
\begin{gather}
    \int_{h}^{L-r_\mathrm{H}} dx (...) = \int_h^R dx (...) + \int_R^{L-r_\mathrm{H}} dx (...),
\end{gather}
by introducing an intermediary radius $R$, where in the first term we keep the divergent parts as $h \to 0$, and in the second term we keep the divergent parts as $L \gg r_H$. The second term is considered to be contribution of the vacuum that surrounds the system at large distances and can be omitted \cite{tHooft:1984kcu}. We identify the most divergent part as $h\to 0$, and equate that term with the Bekenstein-Hawking entropy, $S_{BH}$. This way, we evaluate $h$, and plugging it back into the entropy, we recover the value of the correction term to the Bekenstein-Hawking entropy from the less divergent terms. \\

In what follows we will generalize the method from \cite{Sarkar:2007uz} to the case of charged fields. To expand our consideration to charged black holes and charged probes, we will solve the Klein-Gordon equation using the minimal substitution $\partial_\mu \to \partial_\mu + \frac{i}{\hbar}q A_\mu$, where $A_\mu=(A_0(r),0,0,0)$ describes the 4-vector potential of the point-charged black hole, and $q$ is the charge of the scalar field we will be probing the spacetime with. The Klein-Gordon equation now becomes
\begin{gather}
    \bigg( \frac{1}{\sqrt{-g}}\bigg(\partial_\mu (\sqrt{-g} g^{\mu\nu}\partial_\nu) + \partial_\mu \left(\sqrt{-g} g^{\mu\nu} \frac{i}{\hbar} qA_\nu\right)  + \frac{i}{\hbar} qA_\mu (\sqrt{-g} g^{\mu\nu} \partial_\nu)\bigg) - \frac{q^2 }{\hbar^2} g^{\mu \nu} A_\mu A_\nu - \frac{\mu^2}{\hbar^2}\bigg)\Phi=0.
\end{gather}
Using the same ansatz (\refeq{ansatz}) as before , we arrive at the following radial equation
\begin{equation}
    R''(r)+\left(\frac{W^2(r)}{\hbar^2}-\Delta(r)\right)R(r)=0,
\end{equation}
where we defined a new quantity
\begin{gather}
    W^2(r)=\frac{1}{G^2(r)}\left((E-qA_0(r))^2-g(r)\bigg[\mu^2+ \left(\frac{\ell(\ell+D-1)\hbar^2}{r^2}\right)\bigg]\right),
\end{gather}
for which the radial equation then has the same structure as before (\refeq{V2}), with $V^2(r) \to W^2(r)$, i.e., $E\to E-qA_0(r)$. This supstitution applies to all expression describing charged systems, and as such should also be applied to the condition on $\ell_{\text{max}}$ (\refeq{lmax}). Thus, in order to calculate the entropy up to arbitrary order in the WKB approximation for a charged black hole that is probed with a charged scalar field, we need to replace $E$ with $E-qA_0(r)$ in the defining equations for $N_{2n}(E)$, (\refeq{broj stanja}), and plugging that into the equations for the free energy and entropy (\refeq{F2n S2n}).

\section{Examples and limiting cases}
\subsection{Probing the Reissner-Nordstr\"{o}m black hole with a charged scalar field}
In the case of the Reissner-Nordstr\"{o}m black hole with a charged massless probe, we know the electric potential it generates, as well as its metric
\begin{gather}
    A_0(r)=\frac{Q}{r}, \quad f(r)=g(r)=1-\frac{2M}{r}+\frac{Q^2}{r^2}, \quad D=2, \quad \mu=0, \\\nonumber \quad r_\pm=M\pm\sqrt{M^2-Q^2},  \quad r_\mathrm{H}=r_+.
\end{gather}
We will now calculate the entropy of this black hole up to the second order in the WKB approximation.
\subsubsection{Zeroth order}
In the zeroth order we have
\begin{equation}
P_0(r)=\frac{1}{g(r)}\left[(E - qA_0)^2-g(r) \frac{\hbar^2\ell(\ell+1)}{r^2}\right]^{1/2},
\end{equation}
and after plugging it into $N_0(E)$
\begin{equation}
N_0(E)=\frac{1}{\hbar \pi}\int_{r_++h}^{L} dr \int_{0}^{\ell_{\text{max}}} d\ell (2\ell+1) P_0(r),
\end{equation}
using the substitution $\lambda=\frac{\ell(\ell+1)\hbar^2}{r^2}$, and executing the $d\ell$ integration, we arrive at the radial integral for $N_0(E)$
\begin{equation}
N_0(E)=\frac{2}{3 \pi \hbar^3} \int_{r_{+}+h}^L \frac{\left(E-q\frac{Q}{r}\right)^3r^2}{g(r)^2}dr.
\end{equation}
Substituting $x=r-r_+$, and keeping the dominant near-horizon terms leads to the expression
\begin{equation}
   \begin{split}
   	 N_0(E)&= \frac{2}{3\pi \hbar^3}\frac{r_+^6 \left(E-\frac{qQ}{r_+}\right)^3}{(r_+-r_-)^2}\frac{1}{h} \\
    &+\frac{2}{3\pi \hbar^3}\left(\frac{3 r_+^5\left(2E-\frac{qQ}{r_+}\right)\left(E-\frac{qQ}{r_+}\right)^2}{(r_+-r_-)^2}-\frac{2r_+^6\left(E-\frac{qQ}{r_+}\right)^3}{(r_+-r_-)^3} \right)\ln{\left(\frac{\alpha}{h}\right)}.
   \end{split}
    \label{RNn0}
\end{equation}
The first term has the same structure as the number of modes in \cite{Gupta:2022oel}, while the second is the logarithmic correction. \\

Note that the constant $\alpha$ here signifies a free constant that we can choose however we like, due to the fact that the entropy is defined up to a constant, and the fact that we can omit the vacuum contribution signified by terms that contain $L$. We exploit this fact in integrals of the type
\begin{gather}
    \int_h^{L-r_\mathrm{H}}\frac{1}{x}=\ln\left(\frac{L-r_\mathrm{H}}{h}\right) = \ln\left(\frac{\alpha}{h}\right) + \ln\left(\frac{L-r_\mathrm{H}}{\alpha}\right) \approx \ln\left(\frac{\alpha}{h}\right).
\end{gather}
We choose $\alpha$ in such a way that, after we equate the most divergent part in the entropy as the brick wall cut-off $h\to 0$ with the Bekenstein-Hawking entropy, and after plugging that into the logarithm, the argument of the logarithm has the form $\frac{\mathcal{A}}{\ell^2_{pl}}$. We do this because we can use these types of corrections to investigate the corrections to Planckian sized black holes, as a means to investigate the connection between quantum mechanics and gravity. Since we can redefine the entropy up to a constant, we can add and subtract logarithms until we reach that form. \\

The free energy of (\refeq{RNn0}) is then given by
\begin{align}
    F_0=&-\frac{2}{3\pi \hbar^3}\frac{r_+^6}{(r_+-r_-)^2}\frac{1}{h}K_1(\beta)\\\nonumber &-\frac{2}{3\pi\hbar^3} \left(\frac{3r_+^5}{(r_+-r_-)^2}K_2(\beta)-\frac{2r_+^6}{(r_+-r_-)^3}K_1(\beta)\right)\ln{\left(\frac{\alpha}{h}\right)},
\end{align}
where the functions $K_1(\beta)$, $K_2(\beta)$ are given by
\begin{align}
    K_1(\beta)&=\int_0^\infty dE\frac{\left(E-\frac{qQ}{r_+}\right)^3}{e^{\beta E}-1} \\\nonumber &= \frac{\Gamma(4)\zeta(4)}{\beta^4}-\frac{3qQ\Gamma(3)\zeta(3)}{r_+\beta^3}+\frac{3q^2Q^2 \Gamma(2)\zeta(2)}{r_+^2\beta^2}-\frac{q^3Q^3\Gamma(1)\zeta(1)}{r_+^3\beta},
\end{align}
and
\begin{align}
    K_2(\beta)&=\int_0^\infty \frac{\left(2E-\frac{qQ}{r_+}\right)\left(E-\frac{qQ}{r_+}\right)^2}{e^{\beta E}-1} \\\nonumber &=\frac{2\Gamma(4)\zeta(4)}{\beta^4}-\frac{5qQ\Gamma(3)\zeta(3)}{r_+\beta^3}+\frac{4q^2Q^2 \Gamma(2)\zeta(2)}{r_+^2\beta^2}-\frac{q^3Q^3\Gamma(1)\zeta(1)}{r_+^3\beta}.
\end{align}
The functions $K_1(\beta)$ and $K_2(\beta)$ have an infinite contribution from the electrostatic self-energy of the charge $q$ of the scalar particle, contained in the $\zeta(1)$ term, which we can regulize by rescaling $S_0^{reg}=S_0(\beta)-S_0(\beta= \infty)$. This way, we ensure that $S=0$ when $T=1/\beta=0$, to be in accordance with the third law of thermodyanimcs. Evaluating the entropy at the Hawking temperature (\refeq{Hawkingova temperatura}) we have
\begin{equation}
\begin{split}
	    S_0=&\frac{1}{h}\left(\frac{r_+-r_-}{360}-\frac{3qQr_+\zeta(3)}{4\pi^3\hbar}+\frac{q^2Q^2 r_+^2}{6 (r_+-r_-)\hbar^2}\right)\\
        &+ \left(\frac{2r_+-3r_-}{180 r_+} - \frac{3qQ \zeta(3) (3r_+-5r_-)}{4\pi^3(r_+-r_-)\hbar} +  \frac{q^2Q^2r_+(r_+-2r_-)}{3(r_+-r_-)^2 \hbar^2}\right) \ln{\left(\frac{\alpha}{h}\right)}.
        \label{entropija RN}
\end{split}
\end{equation}
Note that even in the zeroth order of the WKB approximation we obtain logarithmic corrections to the entropy.
\subsubsection{Second order}
Generally, we can decompose $P_2(r)$ as
\begin{gather}
    P_2(r)=\left(\frac{P_2^{(0)}(r)}{\mathcal{G}(\mathcal{E},r)}\right)+\lambda(r)\left(\frac{P_2^{(1)}(r)}{\mathcal{G}^3(\mathcal{E},r)}\right)+\lambda^2(r)\left(\frac{P_2^{(2)}(r)}{\mathcal{G}^5(\mathcal{E},r)}\right),
    \label{P2 dekompozicija}
\end{gather}
where we define
\begin{equation}
\mathcal{G}(\mathcal{E},r)=[\mathcal{E}-\lambda(r)]^{1/2},
\label{G}
\end{equation}
with $\mathcal{E}=\left(E-q\frac{Q}{r}\right)^2$ and
\begin{equation}
\lambda(r)=\ell(\ell+1)\hbar^2\frac{g(r)}{r^2},
\end{equation}
and where $P_2^{(0)}(r),P_2^{(1)}(r),P_2^{(2)}(r)$ are given by 
\begin{equation}
P_2^{(0)}(r)=-\frac{g'(r)}{2r},
\label{P20}
\end{equation}
\begin{equation}
P_2^{(1)}(r)=\frac{3g(r)}{4r^2}-\frac{3g'(r)}{4r}+\frac{g''(r)}{8}+\frac{g'(r)^2}{8g(r)},
\label{P21}
\end{equation}
\begin{equation}
P_2^{(2)}(r)=\frac{5g(r)}{8r^2}-\frac{5g'(r)}{8r}+\frac{5g'(r)^2}{32g(r)}.
\label{P22}
\end{equation}
$N_2(E)$ is defined as
\begin{equation}
N_2(E)=\frac{\hbar}{\pi}\int_{r_{+}+h}^Ldr\int_0^{\ell_{\text{max}}}d\ell (2\ell+1) P_2(r).
\end{equation}
Using the following relations for (\refeq{G})
\begin{gather}
\frac{1}{\mathcal{G}(\mathcal{E},r)}=2\frac{\partial\mathcal{G}(\mathcal{E},r)}{\partial\mathcal{E}},\quad \frac{1}{\mathcal{G}^3(\mathcal{E},r)}=-4\frac{\partial^2\mathcal{G}(\mathcal{E},r)}{\partial\mathcal{E}^2}, \quad \frac{1}{\mathcal{G}^5(\mathcal{E},r)}=\frac{8}{3}\frac{\partial^3\mathcal{G}(\mathcal{E},r)}{\partial\mathcal{E}^3},
\label{relacije za G}
\end{gather}
and plugging (\ref{relacije za G}) into (\refeq{P2 dekompozicija}), we have
\begin{gather}
    \hbar N_2(E)=\frac{1}{\pi}\int_{r_{\mathrm{H}}+h}^L \frac{r^2}{g(r)}dr\int_0^{\mathcal{E}}\Bigg[2\frac{\partial\mathcal{G}(\mathcal{E},\lambda)}{\partial\mathcal{E}}P_2^{(0)}(r)d\lambda -4\lambda\frac{\partial^2\mathcal{G}(\mathcal{E},\lambda)}{\partial\mathcal{E}^2}P_2^{(1)}(r)d\lambda+\frac{8}{3}\lambda^2\frac{\partial^3\mathcal{G}(\mathcal{E},\lambda)}{\partial\mathcal{E}^3}P_2^{(2)}(r)d\lambda\Bigg].
\end{gather}
We now use the Leibniz rule
\begin{gather}
    \frac{\partial}{\partial x}\int_{a(x)}^{b(x)} f[x,t] dt = f[x,b(x)]\left(\frac{db(x)}{dx}\right) - f[x,a(x)]\left(\frac{da(x)}{dx}\right) +\int_{a(x)}^{b(x)}\left[\frac{\partial f(x,t)}{\partial x}\right]dt,
\end{gather}
to extract the divergences that would arise if we were to execute the $\lambda$ integration on the right-hand side. Applying the Leibniz rule to the first term in (\refeq{relacije za G}), with $a(\mathcal{E})=0$ and $b(\mathcal{E})=\mathcal{E}$, we only obtain the finite contribution
\begin{equation}
\int_0^{\mathcal{E}}  P_2^{(0)}(r) \frac{\partial \mathcal{G}(\mathcal{E}, \lambda)}{\partial \mathcal{E}} d \lambda=\frac{\partial}{\partial \mathcal{E}} \int_0^{\mathcal{E}}  P_2^{(0)}(r) \mathcal{G}(\mathcal{E}, \lambda) d\lambda.
\end{equation}
Applying it to the second and third term we have
\begin{align}
    \int_0^{\mathcal{E}} \lambda \frac{\partial^2 \mathcal{G}(\mathcal{E}, \lambda)}{\partial \mathcal{E}^2} d \lambda  
    &= \frac{\partial}{\partial \mathcal{E}} \int_0^{\mathcal{E}} \lambda \frac{\partial \mathcal{G}(\mathcal{E}, \lambda)}{\partial \mathcal{E}} d \lambda-\left.\mathcal{E} \frac{\partial \mathcal{G}(\mathcal{E}, \lambda)}{\partial \mathcal{E}}\right|_{\mathcal{E}=\lambda} 
    \\\nonumber &= \frac{\partial^2}{\partial \mathcal{E}^2} \int_0^{\mathcal{E}} \lambda \mathcal{G}(\mathcal{E}, \lambda) d \lambda-\left.\frac{\mathcal{E}}{2(\mathcal{E}-\lambda)^{1 / 2}}\right|_{\mathcal{E}=\lambda}, \\\nonumber \int_0^{\mathcal{E}} \lambda^2 \frac{\partial^3 \mathcal{G}(\mathcal{E}, \lambda)}{\partial \mathcal{E}^3} d \lambda  &=\frac{\partial^3}{\partial \mathcal{E}^3} \int_0^{\mathcal{E}}  \lambda^2 \mathcal{G}(\mathcal{E}, \lambda) d\lambda  -\left.\left[\frac{\partial}{\partial \mathcal{E}}\left[\frac{\mathcal{E}^2}{2 \mathcal{G}(\mathcal{E}, \lambda)}\right]-\frac{\mathcal{E}^2}{4 \mathcal{G}^3(\mathcal{E}, \lambda)}\right]\right|_{\mathcal{E}=\lambda}.
\end{align}
From the upper two equations, we see that both integrals have a finite and a divergent part. The divergence happens at the turning point $\mathcal{E}=\lambda$. This is a non-physical divergence that arises due to the fact that the WKB approximation is not viable near the turning points of the effective potential \cite{Bender}. 
After discarding the non-physical divergences, $N_2(E)$ is given by
\begin{align}
    \hbar N_2(E)&=\frac{2}{\pi}\int_{r_{+}+h}^L dr\frac{r^2}{g(r)}P_2^{(0)}(r)\frac{\partial}{\partial\mathcal{E}}\int_0^{\mathcal{E}}\mathcal{G}(\mathcal{E},\lambda)d\lambda\\\nonumber&-\frac{4}{\pi}\int_{r_++h}^L dr\frac{r^2}{g(r)}P_2^{(1)}(r)\frac{\partial^2}{\partial\mathcal{E}^2}\int_0^{\mathcal{E}}\lambda\mathcal{G}(\mathcal{E},\lambda)d\lambda\\\nonumber&+\frac{8}{3 \pi}\int_{r_++h}^L dr\frac{r^2}{g(r)}P_2^{(2)}(r)\frac{\partial^3}{\partial\mathcal{E}^3}\int_0^{\mathcal{E}}\lambda^2\mathcal{G}(\mathcal{E},\lambda)d\lambda.
\end{align}
Using the integrals
\begin{gather}
\int_0^{\mathcal{E}}\mathcal{G}(\mathcal{E},\lambda)d\lambda=\frac{2}{3}\mathcal{E}^{3/2}, \quad \int_0^{\mathcal{E}}\lambda\mathcal{G}(\mathcal{E},\lambda)d\lambda=\frac{4}{15}\mathcal{E}^{5/2}, \int_0^{\mathcal{E}}\lambda^2\mathcal{G}(\mathcal{E},\lambda)d\lambda=\frac{16}{105}\mathcal{E}^{7/2},
\end{gather}
$N_2(E)$ is given by
\begin{gather}
    N_2(E)=\frac{1}{\pi \hbar}\int_{r_++h}^L \frac{\left(E-q\frac{Q}{r}\right)r^2}{g(r)}\left(2P_2^{(0)}(r)-4P_2^{(1)}(r)+\frac{16}{3}P_2^{(2)}(r)\right)dr.
\end{gather}
Plugging in (\refeq{P20})-(\refeq{P22}), the final expression for $N_2(E)$ is given by
\begin{gather}
    N_2(E)=\frac{1}{\hbar\pi}\int_{r_++h}^L dr \left(E-q\frac{Q}{r}\right)\bigg[\frac{1}{3}-\frac{4rg'(r)}{3g(r)}+r^2\left(\frac{g'(r)^2}{3g(r)^2}-\frac{g''(r)}{2g(r)}\right)\bigg].
\end{gather}
which leads to the following free energy
\begin{gather}
    F_2=-\frac{1}{\hbar\pi}\int_{r_++h}^Ldr \bigg[\frac{1}{3}-\frac{4rg'(r)}{3g(r)}+r^2\left(\frac{g'(r)^2}{3g(r)^2}-\frac{g''(r)}{2g(r)}\right)\bigg] K_3(\beta)
\end{gather}
with
\begin{gather}
    K_3(\beta)=\int_0^\infty dE \frac{\left(E-\frac{qQ}{r}\right)}{e^{\beta E}-1}= \frac{\Gamma(2)\zeta(2)}{\beta^2}-\frac{qQ \Gamma(1)\zeta(1)}{r\beta}.
\end{gather}
We once again discard the $\zeta(1)$ term, interpreting it as the infinite contribution from the electrostatic self-energy. We are left with
\begin{gather}
    F_2=-\frac{\pi}{6 \hbar \beta^2}\int_{r_++h}^Ldr \bigg[\frac{1}{3}-\frac{4rg'(r)}{3g(r)}+r^2\left(\frac{g'(r)^2}{3g(r)^2}-\frac{g''(r)}{2g(r)}\right)\bigg].
\end{gather}
The entropy evaluated at Hawking temperature is now
\begin{gather}
    S_2 = \frac{\kappa}{6}\int_{r_++h}^Ldr \bigg[\frac{1}{3}-\frac{4rg'(r)}{3g(r)}+r^2\left(\frac{g'(r)^2}{3g(r)^2}-\frac{g''(r)}{2g(r)}\right)\bigg].
\end{gather}
The near-horizon contribution to this integral gives the following expression for the second order entropy
\begin{gather}
    S_2=\frac{r_+-r_-}{36h}-\frac{1}{36}\ln{\left(\frac{\alpha}{h}\right)}.
\end{gather}
\subsubsection{Total entropy up to second order}
The total entropy up to second order is then
\begin{equation}
    \begin{split}
    	S &= S_0 + S_2  \\
    	&=\frac{1}{h}\left(\frac{11(r_+-r_-)}{360}-\frac{3qQr_+\zeta(3)}{4\pi^3\hbar}+\frac{q^2Q^2 r_+^2}{6 (r_+-r_-)\hbar^2}\right)\\
    	 &+ \left(-\frac{(r_++r_-)}{60 r_+} - \frac{3qQ \zeta(3) (3r_+-5r_-)}{4\pi^3(r_+-r_-)\hbar} + \frac{q^2Q^2r_+(r_+-2r_-)}{3(r_+-r_-)^2 \hbar^2}\right) \ln{\left(\frac{\alpha}{h}\right)}.
    \end{split}
    \label{RNS}
\end{equation}
Equating the most divergent part in $h\to 0$ with the Bekenstein-Hawking entropy $S_{BH}=\frac{\mathcal{A}}{4\ell^2_{Pl}}$, as that's the entropy recovered when there's no brick wall, and using $\mathcal{A}=4r_+^2 \pi$, we obtain the value of $h$
\begin{gather}
    h=\frac{\ell_{Pl}^2}{\pi r_+^2}\left(\frac{11(r_+-r_-)}{360}-\frac{3qQr_+\zeta(3)}{4\pi^3\hbar}+\frac{q^2Q^2 r_+^2}{6 (r_+-r_-)\hbar^2}\right).
    \label{RNh}
\end{gather}
Plugging $h$ back into (\refeq{RNS}), and choosing $\alpha$ such that the argument of the logarithm is $\mathcal{A}/\ell^2_{Pl}$, we arrive at
\begin{gather}
    S=S_{BH}+ \left(-\frac{(r_++r_-)}{60 r_+} - \frac{3qQ \zeta(3) (3r_+-5r_-)}{4\pi^3(r_+-r_-)\hbar} + \frac{q^2Q^2r_+(r_+-2r_-)}{3(r_+-r_-)^2 \hbar^2}\right) \ln{\left(\frac{\mathcal{A}}{\ell^2_{Pl}}\right)}.
    \label{RN entropija lPl}
\end{gather}

\subsection{Probing the QBTZ black hole with a charged scalar field}
The electric potential of the QBTZ black hole with a charged massless probe, and its metric are given by
\begin{gather}
    A_0(r)=Q\ln\left(\frac{r}{l}\right), \quad f(r)=g(r)=-M+\frac{r^2}{l^2}-2Q^2\ln\left(\frac{r}{l}\right), \quad r_\mathrm{H}=r_+, \quad D=1, \quad \mu=0.
\end{gather}
Once again, we calculate the entropy up to second order.
\subsubsection{Zeroth order}
The number of states at the zeroth order is defined as
\begin{gather}
    N_0(E)=\frac{2}{\hbar \pi} \int_{r_++h}^L dr \int_0^{\ell_{\text{max}}}d\ell P_0(r),
\end{gather}
with
\begin{gather}
    P_0(r)=\frac{1}{g(r)}\left[\left(E-qQ\ln\left(\frac{r}{l}\right)\right)^2-g(r)\frac{\ell^2\hbar^2}{r^2}\right]^{1/2}.
\end{gather}
After performing the $\ell$ integration, we are left with
\begin{gather}
    N_0(E)=\frac{1}{2\hbar^2}\int_{r_++h}^L dr \frac{r}{g(r)^{3/2}}\left(E-qQ\ln\left(\frac{r}{l}\right)\right)^2.
\end{gather}
The free energy is given by
\begin{gather}
    F_0=-\frac{1}{2\hbar^2}  \int_{r_++h}^L  dr \frac{r}{g(r)^{3/2}}K_4(\beta,r) , \quad K_4(\beta,r)=\int_0^\infty dE \frac{\left(E-qQ\ln{\left(\frac{r}{l}\right)}\right)^2}{e^{\beta E}-1} \\\nonumber K_4(\beta,r)= \frac{\Gamma(3)\zeta(3)}{\beta^3}-\frac{2qQ\ln{\left(\frac{r}{l}\right)} \Gamma(2)\zeta(2)}{\beta^2}+ \frac{q^2Q^2 \ln{\left(\frac{r}{l}\right)}^2 \Gamma(1) \zeta(1)}{\beta}.
\end{gather}
The $\zeta(1)$ term leads to a divergence, and just like in the RN case, we interpret it as the electrostatic self-energy, and we discard it. The entropy has the form
\begin{gather}
    S_0= \frac{1}{2\hbar^2}\int_{r_+ + h}^L dr \left(\frac{6 \zeta(3)}{\beta^2}-4qQ\ln{\left(\frac{r}{l}\right)}\frac{\zeta(2)}{\beta}\right) \frac{r}{g(r)^{3/2}}.
\end{gather}
After integrating, and evaluating the entropy in the near-horizon limit at the Hawking temperature,
\begin{gather}
    T_\mathrm{H}=\frac{1}{\beta}=\frac{\kappa \hbar}{2 \pi} = \frac{g'(r_+)\hbar}{4\pi}, \quad g'(r_+)=\frac{2r_+}{l^2}-\frac{2Q^2}{r_+}
\end{gather}
we have 
\begin{align}
    S_0 = \frac{1}{\sqrt{h}}\left( \frac{3\zeta(3) \sqrt{g'(r_+)} r_+}{8\pi^2} - \frac{Qq\zeta(2) r_+ \ln{\left(\frac{r_+}{l}\right)}}{\pi \sqrt{g'(r_+)} \hbar}\right) + \sqrt{h} \left( \frac{Qq \zeta(2) \ln{\left(\frac{r_+}{l}\right)}}{\pi \sqrt{g'(r_+)} \hbar}+ \frac{qQ\zeta(2)}{\pi \sqrt{g'(r_+)}\hbar}- \frac{3 \zeta(3) \sqrt{g'(r_+)}}{8\pi^2}\right).
    \label{QBTZ entropija}
\end{align}
\subsubsection{Second oder}
In the second order, the number of states has the following form
\begin{equation}
N_2(E)=\frac{2 \hbar}{\pi}\int_{r_++h}^{L} dr \int_{0}^{\ell_{\text{max}}} d\ell P_2(r),
\end{equation}
and after plugging in the general decomposition (\refeq{P2 dekompozicija}) we are left with
\begin{gather}
N_2(E)=\frac{2 \hbar}{\pi}\int_{r_{\mathrm{H}}+h}^{L} dr \int_{0}^{\ell_{\text{max}}} d\ell \left[\left(\frac{P_2^{(0)}(r)}{\mathcal{G}(\mathcal{E},r)}\right)+\lambda(r)\left(\frac{P_2^{(1)}(r)}{\mathcal{G}^3(\mathcal{E},r)}\right)+\lambda(r)^2\left(\frac{P_2^{(2)}(r)}{\mathcal{G}^5(\mathcal{E},r)}\right)\right],
\end{gather}
where
\begin{equation}
    \lambda(r)=\ell^2 \hbar^2 \frac{g(r)}{r^2}, \quad \ell_{\text{max}}=\frac{\sqrt{\mathcal{E}}r}{\hbar \sqrt{g(r)}}, \quad \mathcal{G}=\sqrt{\mathcal{E}-\lambda(r)}, \quad \mathcal{E}=\left(E-qQ\ln\left(\frac{r}{l}\right)\right)^2.
\end{equation}
Using the following integrals (where we already ignored the non-physical WKB divergences)
\begin{gather}
\int_0^{\sqrt{\mathcal{E}}r/(\hbar \sqrt{g})}\frac{d\ell}{\mathcal{G}(\mathcal{E},\ell)}=\frac{\pi r}{2 \sqrt{g(r)}\hbar}, \quad \int_0^{\sqrt{\mathcal{E}}r/(\hbar \sqrt{g})}\frac{\ell^2 \hbar^2 g(r)/r^2 d\ell}{\mathcal{G}^3(\mathcal{E},\ell)}=-\frac{\pi r}{2 \sqrt{g(r)}\hbar}, \quad \int_0^{\sqrt{\mathcal{E}}r/{(\hbar \sqrt{g})}}\frac{\ell^4 \hbar^4 g(r)^2/r^4 d\ell }{\mathcal{G}^5(\mathcal{E},\ell)}=\frac{\pi r}{2 \sqrt{g(r)}\hbar},
\end{gather}
we obtain
\begin{gather}
N_2(E)=\int_{r_++h}^{L} \frac{r dr}{\sqrt{g(r)}} \bigg[P_2^{(0)}(r)- P_2^{(1)}(r) + P_2^{(2)}(r)\bigg].
\end{gather}
The only contribution to $F_2$ is the divergent self-energy,
\begin{gather}
    F_2 = \frac{\zeta(1)\Gamma(1)}{\beta}\int_{r_++h}^{L} \frac{r dr}{\sqrt{g(r)}} \bigg[P_2^{(0)}(r)- P_2^{(1)}(r) + P_2^{(2)}(r)\bigg],
\end{gather}
which leads to a vanishing entropy contribution in the 2nd order of WKB after regularization.\\

The entropy up to the second order is then just the zeroth order entropy, (\refeq{entropija RN}).
\begin{gather}
    S = \frac{1}{\sqrt{h}}\left( \frac{3\zeta(3) \sqrt{g'(r_+)} r_+}{8\pi^2} - \frac{qQ\zeta(2) r_+ \ln{\left(\frac{r_+}{l}\right)}}{\pi \sqrt{g'(r_+)} \hbar}\right) + \sqrt{h} \left( \frac{qQ \zeta(2) \ln{\left(\frac{r_+}{l}\right)}}{\pi \sqrt{g'(r_+)} \hbar}+ \frac{qQ\zeta(2)}{\pi \sqrt{g'(r_+)}\hbar}- \frac{3 \zeta(3) \sqrt{g'(r_+)}}{8\pi^2}\right).
\end{gather}
Repeating the above procedure of equating the most divergent part in $h$ with the Bekenstein-Hawking entropy, $S=\frac{\mathcal{A}}{4\ell^2_{Pl}}$, with $\mathcal{A}=2\pi r_+$, $\sqrt{h}$ is given by
\begin{gather}
    \sqrt{h}=\frac{2\ell^2_{Pl}}{\pi}\left( \frac{3\zeta(3) \sqrt{g'(r_+)}}{8\pi^2} - \frac{Qq\zeta(2) \ln{\left(\frac{r_+}{l}\right)}}{\pi \sqrt{g'(r_+)} \hbar}\right),
    \label{hQBTZ}
\end{gather}
and the entropy is given by
\begin{gather}
    S=S_{BH}+ G + H\ln\left(\frac{\mathcal{A}}{2\pi l}\right)+I\ln^2\left(\frac{\mathcal{A}}{2\pi l}\right),
    \label{SBTZ simbolicki}
\end{gather}
with
\begin{gather}
    G=\frac{3 Q \ell_{Pl}^{2} q \zeta{\left(2 \right)} \zeta{\left(3 \right)}}{4 \pi^{4} \hbar} - \frac{9 g'(r_{H}) \ell_{Pl}^{2} \zeta^{2}{\left(3 \right)}}{32 \pi^{5}}\;,\\
    H=- \frac{2 Q^{2} \ell_{Pl}^{2} q^{2} \zeta^{2}{\left(2 \right)}}{\pi^{3} g'(r_{H}) \hbar^{2}} + \frac{3 Q \ell_{Pl}^{2} q \zeta{\left(2 \right)} \zeta{\left(3 \right)}}{2 \pi^{4} \hbar}\;,\\
    I=- \frac{2 Q^{2} \ell_{Pl}^{2} q^{2} \zeta^{2}{\left(2 \right)}}{\pi^{3} g'(r_{H}) \hbar^{2}}\;.
\end{gather}
In this case we see that the corrections are in general higher powers of logarithms of the black hole area.
\subsection{Simultaneous $Q \to 0$ and $q\to 0$ limit}
In this subsection we will look at the limits of $Q \to 0$ and $q\to 0$ of the previous two subsections.
\subsubsection{Schwarzschild black hole}
In the case of the Reissner-Nordstr\"{o}m black hole, the limits $Q\to 0$ and $q\to 0$ lead to the Schwarzschild black hole. The entropy for the Schwarzschild black hole up to the second WKB order is then given by applying these limits to (\refeq{RN entropija lPl}) as
\begin{gather}
    S=S_{BH}-\frac{1}{60}\ln\left(\frac{\mathcal{A}}{\ell^2_{Pl}}\right),
    \label{S SCH}
\end{gather}
which coincides with \cite{Sarkar:2007uz}. Notice that the factor in front of the logarithm is a constant.
\subsubsection{BTZ black hole}
In the case of the QBTZ black hole, the limits $Q\to 0$ and $q\to 0$ lead to the BTZ black hole. Applying these limits to (\refeq{SBTZ simbolicki}) and using $\mathcal{A}=2\pi r_+$, $r_+=l\sqrt{M}$, $g(r)=\frac{1}{l^2}(r^2-r_\mathrm{H}^2)$ and $g'(r)=\frac{2r}{l^2}$, the entropy for the BTZ black hole up to the second WKB order is then given by
\begin{gather}
    S=S_{BH}-\frac{18 \zeta^2(3)}{91\pi^6}\frac{\ell^2_{Pl}}{l^2} \mathcal{A}.
    \label{SBTZ}
\end{gather}
Notice that the correction is linear in $\mathcal{A}$.
\subsection{$q\to 0$ limit}
In the following subsection we will look at the behaviour of a neutral scalar probe near a charged black hole.
\subsubsection{Reissner-Nordstr\"{o}m black hole}
The entropy of the RN black hole probed by a neutral probe is given by
\begin{gather}
    S=S_{BH}-\frac{(r_+-r_-)}{60r_+}\ln\left(\frac{\mathcal{A}}{\ell^2_{Pl}}\right).
\end{gather}
Plugging in the definitions of $r_\pm$, the entropy is given by
\begin{gather}  
    S=S_{BH}-\frac{\sqrt{M^2-Q^2}}{30\left(M+\sqrt{M^2-Q^2}\right)}\ln\left(\frac{\mathcal{A}}{\ell^2_{Pl}}\right).
    \label{RN entropija lPl sredjena}
\end{gather}
\subsubsection{QBTZ black hole}
The metric function for the QBTZ black hole is given by
\begin{gather}
    g(r)=-M+\frac{r^2}{l^2}-2Q^2 \ln\left(\frac{r}{l}\right).
\end{gather}
The only difference from the entropy in (\refeq{SBTZ}) comes from the charge term in $g'(r_+)$. The entropy is given by
\begin{gather}
    S=S_{BH}-\frac{18 \zeta^2(3)}{91\pi^6}\frac{\ell^2_{Pl}}{l^2}\mathcal{A}\left(1-\frac{4\pi^2 Q^2 l^2}{\mathcal{A}^2}\right).
\end{gather}
\section{Noncommutative generalization}
In this Section we will derive the modified radial equations of motion for the noncommutative charged black hole and charged scalar field in any number of dimensions and then we will consider the entropy for such black holes. \\

As explained in the Introduction, an established approach of defining field theory on noncommutative spacetime is to take a commutative field theory and deform its symmetry using the twist approach \cite{Gupta:2022oel, Juric:2022bnm} which necessarily propagates to deforming the pointwise product of spacetime functions. We shall consider $U(1)$ scalar field theories on background charged spherical black hole geometries (\refeq{metrika}) in $D+2$ dimensions
\begin{equation}
	S[\Phi] = \int   \sqrt{-g}\left[g^{\mu\nu} D_\mu \Phi D_\nu \Phi - \frac{m^2}{2}\Phi^\dagger\Phi\right]\;,
\end{equation}
and deform their diffeomorphism and $U(1)$ symmetry. Our noncommutative space will be defined via the angular twist\footnote{Since $\partial_t$ and $\partial_\phi$ Killing vector fields commute, the angular twist is both Abelian and Killing. }
\begin{equation}
	\mathcal{F} = \exp\Big[-\frac{i}{2}\Theta^{\mu\nu}\partial_\mu\otimes\partial_\nu \Big] = \exp\Big[-i \frac{a}{2}\left(\partial_t \otimes \partial_\phi - \partial_\phi \otimes \partial_t\right)\Big]\;,
\end{equation}
with $a$ being the noncommutativity paremeter dictating the strength of spacetime noncommutativity. For the choice of spherically symmetric charged BH metrics, the quantization of pointwise spacetime algebra does not affect the components of the metrics\footnote{Even though the diffeormophism symmetry underlying General Relativity is deformed, the spherically symmetric static black hole solutions to noncommutative field equations are the same as undeformed ones. It is important to note that some geometric properties do get modified, such as the geodesic motion and quasinormal modes, but those do not affect our calculation since we are considering the deformed $U(1)$ scalar theory in which gravity only plays the role of a metric background field.}, as the twist is constructed from Killing vectors of the  metric. Therefore, the deformed, noncommutative $U(1)_\star$ scalar field action looks as follows
\begin{equation}
	S [\hat{\Phi}] = \int \sqrt{-g}\left[g^{\mu\nu}\left( \hat{D}_\mu \hat{\Phi}\right)^+ \star \left( \hat{D}_\nu \hat{\Phi}\right) - \frac{m^2}{\hbar^2} \hat{\Phi}^+\star \hat{\Phi} \right]\;,
\end{equation} 
with the $U(1)_\star$ covariant derivative being
\begin{equation}
	\hat{D}_\mu \hat{\Phi} = \partial_\mu\hat{\Phi} - i\frac{q}{\hbar}\hat{A}_\mu\star\hat{\Phi}\;.
\end{equation}
The described procedure will have a twofold effect on the equation of motion for the scalar field. Firstly, commutative products are replaced with star products (which are covariant with the twisted Hopf algebra action)
\begin{equation}
	\hat{f}\star\hat{g} = \cdot\left(\mathcal{F}^{-1}\hat{f}\otimes \hat{g}\right) = f\cdot g + \frac{ia}{2}\left(\partial_t \hat{f} \cdot\partial_\phi \hat{g} - \partial_\phi \hat{f}\cdot \partial_t \hat{g}\right) + \mathcal{O}(a^2)\;,
\end{equation}
and secondly, since the gauge symmetry sector also got deformed, it is necessary to expand noncommutative fields themselves in terms of their commutative gauge counterparts, i.e., do the Seiberg-Witten expansion \cite{Seiberg:1999vs}
\begin{equation}
\begin{split}
		\hat{\Phi} &= \Phi - \frac{q}{4\hbar}\Theta^{\rho\sigma}A_\rho\left(\partial_\sigma \Phi + \left(\partial_\sigma -i\frac{q}{\hbar}A_\sigma\right)\Phi \right) + \mathcal{O}(a^2)\\
		\hat{A}_\mu &= A_\mu - \frac{q}{2\hbar}\Theta^{\rho\sigma}A_\rho\left(\partial_\sigma A_\mu + F_{\sigma\mu}\right) + \mathcal{O}(a^2)\;,
\end{split}
\end{equation}
which adds even more additional terms to the action. With that said, expanding the Seiberg-Witten map and star products, we are left with the action \cite{Ciric:2017rnf}
\begin{equation}
\begin{split}
		S[\Phi] =&\int\sqrt{-g}\left[ g^{\mu\nu}D_\mu\Phi^\dagger D_\nu\Phi - \frac{m^2}{\hbar^2}\Phi^\dagger\Phi + \frac{qm^2}{2\hbar^3}\Theta^{\alpha\beta}F_{\alpha\beta}\Phi^\dagger\Phi\right.\\ &+\left.\frac{q}{2\hbar}\Theta^{\alpha\beta}\left(-\frac{1}{2}D_\mu\Phi^\dagger F_{\alpha\beta}D_\nu \Phi+ D_\mu\Phi^\dagger F_{\alpha\nu}D_\beta \Phi + D_\beta\Phi^\dagger F_{\alpha\mu}D_\nu\Phi\right)\right] + \mathcal{O}(a^2)\;,
\end{split}
\end{equation}
where the Faraday tensor $F_{\mu\nu}$ comes from the point-charge electromagnetic field.
We now vary the expanded action with respect to $\Phi^\dagger$ and obtain the equation of motion for $\Phi$
\begin{equation}
	\begin{split}
		&g^{\mu\nu}\left[\left(\partial_\mu - i\frac{q}{\hbar}A_\mu\right)  D_\nu\Phi - \Gamma^\lambda_{\mu\nu}D_\lambda\Phi\right] - \frac{m^2}{\hbar^2}\Phi + \frac{qm^2}{2\hbar^3}\Theta^{\alpha\beta}F_{\alpha\beta}\Phi\\
		&-\frac{q}{4\hbar}g^{\mu\nu}\Theta^{\alpha\beta}\left[\left(\partial_\mu -  i\frac{q}{\hbar}A_\mu\right)\left(F_{\alpha\beta}D_\nu \Phi\right) -\Gamma^\lambda_{\mu\nu}F_{\alpha\beta}D_\lambda\Phi -2\left(\partial_\mu - i\frac{q}{\hbar}A_\mu\right)\left(F_{\alpha\nu}D_\beta\Phi\right)  +2\Gamma^\lambda_{\mu\nu}F_{\alpha\lambda}D_{\beta}\Phi -2\left(\partial_\beta - i\frac{q}{\hbar}A_\beta\right)\left(F_{\alpha\mu}D_\nu\Phi\right) \right] + \mathcal{O}(a^2)=0\,,
	\end{split}
\label{NCKG eq}
\end{equation}
which, in this formalism, is the NC Klein-Gordon equation and the terms proportional to $\Theta$ are the corrections to the commutative Klen-Gordon. For the sake of generalizing radial equations of motion from \cite{Gupta:2022oel, Juric:2022bnm}, we shall do an intermediary separation of the equation of motion before substituting further to the form of (\refeq{ansatz})
\begin{equation}
	\Phi =R_{\ell m_i}(r) Y_{\ell m_i}(\Omega)e^{-i\frac{Et}{\hbar}}\;.
	\label{Separacija 107}
\end{equation}
The equation (\refeq{NCKG eq}) using (\refeq{Separacija 107}) separates, up to linear order in noncommutativity parameter $a$, into
\begin{equation}
\begin{split}
		&g(r)R''_{\ell m_i}(r) + \left[\frac{Dg(r)}{r} + \frac{g(r)\frac{d}{dr}G(r)}{G(r)}\right]R_{\ell m_i}'(r) + \frac{1}{g(r)}\left[\frac{(E - qA_0(r))^2}{\hbar^2} - g(r)\left(\frac{\mu^2}{\hbar^2} + \frac{\ell(\ell+D-1)}{r^2}\hbar^2\right)\right]R_{\ell m_i}(r) - \\
		\\ &a \frac{imq}{2\hbar}g(r)\left[\frac{D\frac{d}{d r} A_{0}{\left(r \right)}}{r} +\frac{ \frac{d}{d r} A_{0}{\left(r \right)} \frac{d}{d r} G{\left(r \right)}}{ G{\left(r \right)}} + \frac{d^{2}}{d r^{2}} A_{0}{\left(r \right)} + 2\frac{d}{d r} A_{0}{\left(r \right)} \frac{d}{d r} \right]R_{\ell m_i}{\left(r \right)}= 0\;,
\label{NCKG R eq}
\end{split}
\end{equation}
where $m$ is the azimuthal wave number of the hyperspherical harmonic $Y_{lm_i}(\Omega)$. Equation (\refeq{NCKG R eq}) immediately reproduces results from \cite{Gupta:2022oel, Juric:2022bnm} when considering the special cases of RN:
\begin{gather}
    g(r) R''_{\ell m}(r) + \frac{2}{r}\left(1-\frac{GM}{r}\right) R'_{\ell m}(r)-\left[\frac{\ell(\ell+1)}{r^2}-\frac{1}{g(r)}\left(E-\frac{qQ}{r}\right)^2\right]R_{\ell m}(r)-ima\frac{qQ}{r^3}\left[\left(\frac{GM}{r}-\frac{GQ^2}{r^2}\right)R_{\ell m}(r)+rg(r)R'_{\ell m}(r)\right]=0\;,
    \label{NCRNEOM}
\end{gather}
and charged BTZ spacetimes:
\begin{gather}
    R''_m(r)+\frac{1}{g(r)}\left[\frac{\left(E-qQ\ln\left(\frac{r}{l}\right)\right)^2}{g(r)\hbar^2}-\frac{m^2}{r^2}-\frac{\mu^2}{\hbar^2}\right]R_m(r) + \frac{2}{rg(r)}\left(\frac{r^2}{\ell^2}-Q^2+\frac{g(r)}{2}\right)R'_m(r) + im\frac{aqQ}{\hbar r^2 g(r)}\left[rg(r)\frac{d}{dr}+r^2\left(\frac{1}{l^2}-\frac{Q^2}{r^2}\right)\right]R_m(r)=0\;.
    \label{NCQBTZEOM}
\end{gather}
Finally, in order to fully generalize the results (\refeq{V2}), (\refeq{Delta}), we further use the substitution (\refeq{ansatz})
\begin{equation}
	R_{\ell m}(r) = \frac{\psi_{\ell m}(r)}{r^{D/2}\sqrt{G(r)}}\;,
\end{equation}
and by doing so, obtain the following general (up to first order in $a$) radial  NCKG equation for charged probes in charged spherically symmetric black hole spacetime
\begin{equation}
	\psi''_{\ell m_i}(r) + \left(\frac{A(r)}{\hbar} + B(r)\right)\psi'_{\ell m_i}(r) + \left(\frac{W^2(r)}{\hbar^2} + \frac{C(r)}{\hbar} - \Delta(r)\right)\psi_{\ell m_i}(r) = 0
	\label{Korigirana KG}
\end{equation}
with
\begin{equation}
	\begin{split}
		W^2(r) &= \frac{1}{G(r)^2}\left((E-qA_0(r))^2-g(r)\left[\mu^2+ \left(\frac{\ell(\ell+D-1)\hbar^2}{r^2}\right)\right]\right) \\
		\Delta(r) &= \frac{D(D-2)}{4 r^2} + \frac{D}{2 r}\frac{G'(r)}{G(r)} - \frac{(G'(r))^2}{4 G(r)^2} + \frac{G''(r)}{2 G(r)}\\
		A(r) &= -{i a m q \frac{d}{d r} A_{0}{\left(r \right)}}
		\\
		B(r) &= 0 \\
		C(r) &=  - \frac{i a m q \frac{d^{2}}{d r^{2}} A_{0}{\left(r \right)}}{2}
		\;.
	\end{split}
\label{NC koeficijenti}
\end{equation}
We can see in (\refeq{NC koeficijenti}) that every term in the correction to the KG equation is dependent on the electromagnetic field of the black hole and on the charge of the probe, which is in agreement with our previous claims that the twist approach in our considerations only expresses itself in the $U(1)$ part of the action. \\

After finding the coefficients (\refeq{NC koeficijenti}) of generalized radial equation (\refeq{Korigirana KG}), we can extend the brick wall formalism to general equations of the form (\refeq{Korigirana KG}). Let us apply the WKB approximation
\begin{equation}
	\psi_{\ell m_i}(r) = \frac{1}{\sqrt{P_{\ell m_i}(r)}}\exp\left[\frac{i}{\hbar}\int^r P_{\ell m_i}(r')dr'\right]
\end{equation}
to the equation (\refeq{NC koeficijenti}). In doing so, we find the differential equation for $P_{\ell m_i}(r)$
\begin{equation}
	\left(W^2(r) - P_{\ell m_i}(r)^2 + iP_{\ell m_i}(r)A(r)\right) + \hbar\left(C(r) + iP_{\ell m_i}(r)B(r) - \frac{P'_{\ell m_i}(r)}{2P_{\ell m_i}(r)}A(r)\right) = \hbar^2 \left(\frac{1}{2}\frac{P_{\ell m_i}''(r)}{P_{\ell m_i}(r)} - \frac{3}{4}\left(\frac{P_{\ell m_i}'(r)}{P_{\ell m_i}(r)}\right)^2 + \Delta(r) + \frac{P_{\ell m_i}(r)}{2P_{\ell m_i}(r)}B(r)\right)\;,
\end{equation}
in which we can expand $P_{\ell m_i}(r)$ in powers of $\hbar$
\begin{equation}
	P_{\ell m_i}(r) = \sum_{n=0}^\infty \hbar^n P_{n}(r)\;,
	\label{NC P razvoj}
\end{equation}
where we have (for the sake of notation) suppressed the index $l$ and the (partially ordered) multi-index $m_i$, and solve order by order. The solutions up to $n=1$ are given below
\begin{equation}
	\begin{split}
		P_0(r) &= i\frac{A(r)}{2}\pm \sqrt{W^2(r)-\frac{A(r)^2}{4}}\\ 
		P_1(r) &= \frac{A(r)\frac{P'_0(r)}{2P_0(r)}-iB(r)P_0(r)-C(r)}{iA(r)-2P_0(r)}\;.
	\end{split}
\label{P0P1}
\end{equation}
We shall only consider the $+$ branch of the $P_0(r)$ solution in (\refeq{P0P1}). It is, once again, worth noting that all of the equations for $P_n(r)$ are algebraically dependent on $A(r),B(r),C(r),W^2(r),\Delta(r)$ and on derivatives of $P_0(r)$. Focusing on the lowest order of WKB, we can expand $P_0(r)$ for small\footnote{Which is a valid assumption, as $A\sim a$ in our model. As explained before, $a$ is the parameter of spacetime noncommutativity, which is theorized to be at most of the order of Planck length $\ell_{Pl}$ \cite{Balachandran:2009mq}, or smaller, with current upper bound assessments supporting this claim \cite{Carlson:2001sw}.} $A(r)$
\begin{equation}
	P_0 = \sqrt{W^2(r)} + i\frac{A(r)}{2} - \frac{A(r)^2}{8{\sqrt{W^2(r)}}}\;.
	 \label{a^2 razvoj P0}
\end{equation}
We can see that in (\refeq{a^2 razvoj P0}), the dominant term is the commutative contribution, i.e., the charged variant of $P_0$ from (\refeq{P komutativni}), while the remaining two terms are NC corrections. In the expansion (\refeq{NC P razvoj}), the $n$-th contribution to the number of states $N(E)$ arises from the Bohr-Sommerfeld quantization rule
\begin{equation}
N_n(E) = \frac{\hbar^{n-1}}{\pi}\sum_{0\leq |m|\leq m_2\leq...\leq m_{D-1}\leq l \leq l_{\text{max}}} \int_{r_H + h}^L P_{n,lm_i}(r) dr\;.
\end{equation}
Finally, for the noncommutative charged spacetime and charged scalar field, the zeroth noncommutative contribution to the number of states $N(E)$ can be written as the integral
\begin{equation}
	N(E)_{0,\text{NC}} = -\frac{1}{8\hbar} \int_{r_H + h}^L dr\int_0^{l_{\text{max}}(E,r)}dl \sum_{m_{D-1}=0}^{l}...\sum_{m_2=0}^{m_3}\sum_{m=-m_2}^{m_2} \frac{A(r)^2}{{\sqrt{W^2(r)}}}\;,
\label{NC N(E)}
\end{equation}
where $\ell_{\text{max}}$ is the classical restriction on the realness of $P_{\ell m_i}(r)$, whose implicit equation was given in (\refeq{lmax}) (of course, taking into account the minimal substitution $E\rightarrow E-qA_0$). It is important to comment that in (\refeq{NC N(E)}), the $i\frac{A(r)}{2}$ term does not contribute to $N(E)$ as $A(r)$ is linear in $m$ according to (\refeq{NC koeficijenti}), and the sum goes over positive and negative values of $m$ symmetrically. After calculating (\refeq{NC N(E)}), we can use it to calculate the NC correction to entropy using (\refeq{F2n S2n}) as usual.
\subsection{Noncommutative Reissner-Nordstr\"{o}m with a charged scalar field}
Applying the steps explained in Section IV to the Reissner-Nordstr\"{o}m with a charged scalar field and twisting the symmetry with the angular twist, i.e., plugging in the Reissner-Nordstr\"{o}m's data into (\refeq{NC N(E)}), we obtain the noncommutative contribution to the number of states in the lowest order of WKB
\begin{gather}
    N_{0,NC}(E)=\frac{Q^2q^2a^2}{18\hbar^5 \pi}\int_{r_++h}^L dr \left(E-\frac{qQ}{r}\right)^3 \frac{1}{g(r)}.
\end{gather}
The free energy is then
\begin{gather}
    F_{0,NC}=-\frac{Q^2q^2a^2}{18\hbar^5\pi}\int_{r_+ + h}^L dr \frac{1}{g(r)}\left(\frac{\Gamma(4)\zeta(4)}{\beta^4}-3\frac{\Gamma(3)\zeta(3)}{\beta^3}\frac{Qq}{r}+3\frac{\Gamma(2)\zeta(2)}{\beta^2}\frac{Q^2 q^2}{r^2}\right),
\end{gather}
and the entropy
\begin{gather}
    S_{0,NC}=\frac{Q^2q^2a^2}{18\hbar^5\pi}\int_{r_+ + h}^L dr \frac{1}{g(r)}\left(4\frac{\Gamma(4)\zeta(4)}{\beta^3}-9\frac{\Gamma(3)\zeta(3)}{\beta^2}\frac{Qq}{r}+6\frac{\Gamma(2)\zeta(2)}{\beta}\frac{Q^2 q^2}{r^2}\right).
    \label{NC dio NCRN entropije}
\end{gather}
Using
\begin{gather}
    g(r)=\frac{(r-r_+)(r-r_-)}{r^2}, \quad T_{\mathrm{H}}=\frac{1}{\beta}=\frac{\hbar g'(r_+)}{4\pi}, \quad g'(r_+)=\frac{r_+-r_-}{r_+^2},
\end{gather}
The noncommutative entropy in the zeroth order has the form
\begin{gather}
    S_{0,NC}=\frac{Q^2 q^2 a^2 g'(r_+)}{48 \hbar^4 \pi^4}\left(\hbar^2\zeta(4)g'(r_+)-3\pi\hbar \frac{Qq\zeta(3)}{r_+}+4 \pi^2 \frac{Q^2 q^2\zeta(2)}{r_+-r_-}\right)\ln\left(\frac{\alpha}{h}\right).
\end{gather}
The total entropy in the zeroth order for the noncommutative NCRN is then the combination of (\refeq{NC dio NCRN entropije}) and (\refeq{entropija RN}).
\begin{align}
    S_0=&\frac{1}{h}\left(\frac{r_+-r_-}{360}-\frac{3qQr_+\zeta(3)}{4\pi^3\hbar}+\frac{q^2Q^2 r_+^2}{6 (r_+-r_-)\hbar^2}\right)\\\nonumber &+ \left[\frac{2r_+-3r_-}{180 r_+} - \frac{3qQ \zeta(3) (3r_+-5r_-)}{4\pi^3(r_+-r_-)\hbar} +  \frac{q^2Q^2r_+(r_+-2r_-)}{3(r_+-r_-)^2 \hbar^2}\right. \\\nonumber &\left. + \frac{Q^2 q^2 a^2 g'(r_+)}{48 \hbar^4 \pi^4}\left(\hbar^2\zeta(4)g'(r_+)-3\pi\hbar \frac{Qq\zeta(3)}{r_+}+4 \pi^2 \frac{Q^2 q^2\zeta(2)}{r_+-r_-}\right)\right]\ln\left(\frac{\alpha}{h}\right).
\end{align}
$h$ is given by
\begin{gather}
    h=\frac{\ell^2_{Pl}}{r_+^2\pi}\left(\frac{r_+-r_-}{360}-\frac{3qQr_+\zeta(3)}{4\pi^3\hbar}+\frac{q^2Q^2 r_+^2}{6 (r_+-r_-)\hbar^2}\right),
\end{gather}
and the entropy is given by
\begin{align}
    S_0=S_{BH} &+ \Bigg(\frac{2r_+-3r_-}{180 r_+} - \frac{3qQ \zeta(3) (3r_+-5r_-)}{4\pi^3(r_+-r_-)\hbar} +  \frac{q^2Q^2r_+(r_+-2r_-)}{3(r_+-r_-)^2 \hbar^2} \\\nonumber &+ \frac{Q^2 q^2 a^2 g'(r_+)}{48 \hbar^4 \pi^4}\left(\hbar^2\zeta(4)g'(r_+)-3\pi\hbar \frac{Qq\zeta(3)}{r_+}+4 \pi^2 \frac{Q^2 q^2\zeta(2)}{r_+-r_-}\right)\ln\left(\frac{\alpha}{h}\right)\; .
\end{align}
\subsection{Noncommutative QBTZ with a charged scalar field}
Using the equation (\refeq{NC N(E)}) on the case of NCQBTZ, we get the following noncommutative correction to the number of states in the lowest order after perfoming the azimuthal number integration
\begin{gather}
    N_0(E)=\frac{1}{2\hbar^2}\int_{r_++h}^Ldr \left[E^2-2EQq\ln\left(\frac{r}{l}\right)\right]\frac{r}{\sqrt{g(r)}}\left(\frac{1}{g(r)}+\frac{a^2q^2Q^2}{8\hbar^2}\right)
\end{gather}
The entropy is given by
\begin{gather}
    S_0=\frac{1}{2\hbar^2}\int_{r_++h}^L dr \left(\frac{6\zeta(3)}{\beta^2}-4Qq\ln\left(\frac{r}{l}\right)\frac{\zeta(2)}{\beta}\right) \frac{r}{\sqrt{g(r)}}\left(\frac{1}{g(r)}+\frac{a^2q^2Q^2}{8\hbar^2}\right)
\end{gather}
The first term reproduces the commutative part, (\refeq{SBTZ simbolicki}). The noncommutative part after using the near-horizon substitution $x=r-r_+$, aswell as $\ln\frac{r_++x}{l}=\ln\frac{r_+}{l}+\frac{x}{r_+}$, up to $\sqrt{h}$ order is given by
\begin{gather}
    S_{0,NC}=\sqrt{h}\left[\frac{Q^3\zeta(2) a^2 \sqrt{g'(r_+)} q^3 r_+ \ln\left(\frac{r_+}{l}\right)}{8\pi \hbar^3}-\frac{3Q^2 \zeta(3) a^2 (g'(r_+))^{3/2}q^2 r_+}{64\pi^2 \hbar^2}\right].
    \label{NC dio NCQBTZ entropije}
\end{gather}
Notice that the zeroth order noncommutative entropy is of order $\sqrt{h}$, that is, the noncommutative part already serves as a correction to the commutative part, where the most divergent part is of order $1/\sqrt{h}$. The total entropy in the zeroth order for the noncommutative QBTZ is then the combination of (\refeq{NC dio NCQBTZ entropije}) and (\refeq{QBTZ entropija})
\begin{align}
    S &= \frac{1}{\sqrt{h}}\left( \frac{3\zeta(3) \sqrt{g'(r_+)} r_+}{8\pi^2} - \frac{Qq\zeta(2) r_+ \ln{\left(\frac{r_+}{l}\right)}}{\pi \sqrt{g'(r_+)} \hbar}\right) \\\nonumber &+ \sqrt{h} \left( \frac{Q^3\zeta(2) a^2 \sqrt{g'(r_+)} q^3 r_+ \ln\left(\frac{r_+}{l}\right)}{8\pi \hbar^3}-\frac{3Q^2 \zeta(3) a^2 (g'(r_+))^{3/2}q^2 r_+}{64\pi^2 \hbar^2}+  \frac{qQ \zeta(2) \left(\ln{\left(\frac{r_+}{l}\right)}+1\right)}{\pi \sqrt{g'(r_+)} \hbar}- \frac{3 \zeta(3) \sqrt{g'(r_+)}}{8\pi^2}\right).
\end{align}
Equating the most divergent part with $S_{BH}$, we obtain the same $\sqrt{h}$ as in the commutative case (\refeq{hQBTZ}). The total entropy is then
\begin{gather}
    S=S_{BH}+ \sum_{n=0}^2 G_n \ln^n\left(\frac{\mathcal{A}}{2\pi l}\right) + \frac{a^2 Q^2q^2}{\hbar^2} \sum_{n=0}^{2}H_n \ln^n\left(\frac{\mathcal{A}}{2\pi l}\right),
\end{gather}
with
\begin{equation}
\begin{split}
	G_0 = &\frac{3\ell_{Pl}^2 Q \zeta(2) \zeta(3) q}{4 \pi^{4} \hbar} - \frac{9\ell_{Pl}^2 \zeta^2(3) g'(r_{+})}{32 \pi^{5}}\;,\quad G_1 = -\frac{2 \ell_{Pl}^2 Q^{2} \zeta^2(2) q^{2}}{\pi^{3} g'(r_{+}) \hbar^{2}} + 
	\frac{3 \ell_{Pl}^2 Q \zeta(2) \zeta(3) q }{2 \pi^{4} \hbar}\\
	G_2 =& -\frac{2 \ell_{Pl}^2 Q^{2} \zeta^2(2) q^{2}}{\pi^{3} g'(r_{+}) \hbar^{2}}\;,\quad H_0 = - \frac{9 \ell_{Pl}^2  \zeta^2(3) g'(r_{+})^{2} r_{H}}{256 \pi^{5}} \\
	H_1 =& \frac{3\ell_{Pl}^2  Q^{} \zeta(2) \zeta(3) g'(r_{+}) q^{} r_{+}}{16 \pi^{4} \hbar^{}}\;,\quad H_2 = - \frac{\ell_{Pl}^2 Q^{2} \zeta^2(2) q^{2} r_{+}}{4 \pi^{3} \hbar^{2}}\\
\end{split}
\end{equation}
Notice that the entropy now has general corrections of the form $\mathcal{R}(\mathcal{A}) \ln^{m}\left(\mathcal{A}\right)$, with $\mathcal{R}(\mathcal{A})$ being rational functions in black hole area.

\section{Final remarks}

In this paper we have generalized the brick wall method to include charged black holes and charged probes. We applied this method to calculate the entropy for the Reissner-Nordstr\"{o}m and QBTZ black hole. We looked at the Schwarzschild and BTZ limits of the above results, as well as the limits where the probe is neutral. In the Schwarzschild and Reissner-Nordstr\"{o}m case, we obtained corrections to the entropy that are logarithmic in nature, while in the QBTZ and BTZ case, the corrections are rational functions multiplying powers of logarithms in area.\\

Further increasing the generality of our considerations, by using the angular twist we considered the noncommutative action of a charged scalar field in charged spacetime. We have derived the radial noncommutative Klein-Gordon equation in any $D+2$ dimensions, and generalized the WKB method to radial equations of the form arising in this case. As special cases, we have applied the method to $D=2$, the noncommutative Reissner-Nordstr\"{o}m (NCRN) and $D=1$, the noncommutative QBTZ (NCQBTZ). The corrections to entropy are found to be logarithmic in nature for NCRN, while in the NCQBTZ case we encounter corrections of the form $\mathcal{R}(\mathcal{A}) \ln^m\left(\mathcal{A}\right)$. Additionally, the expressions for the entropy in the noncommutative regime are always of the form of entropy in the commutative case with corrections proportional to (powers of) the noncommutativity parameter $a$. This means that the commutative limit $a\rightarrow 0$ is well defined and reproduces the first part of this paper. Finally, we found that for the angular twist $\partial_t \wedge\partial_\phi$, the noncommutativity can affect the entropy only for charged probes in charged spacetimes. That is, because the Hopf algebra deformation of gravity is trivial for the angular twist, since the angular twist is entirely made of commuting Killing vectors of the charged spacetime.\\

Throughout this paper, it may seem like the brick wall is dependent on the mass of the black hole (\refeq{hQBTZ}), (\refeq{S SCH}), (\refeq{RN entropija lPl sredjena}), etc. However this is just a coordinate artifact that is removed once we rewrite it in terms of the coordiante-invariant distance (\refeq{hc}), which shows that the brick-wall can be seen as a property of the horizon, independent of the particular size of a given black hole. An argument could be made that we could have written all the entropies in terms of $h_c$ to make the entropy invariant, by inverting (\refeq{hc}), however that was not necessary. By equating the terms that diverge the most in $h_c$ with the Bekenstein-Hawking entropy and obtaining $h_c$, after plugging it into the entropy, all the additional constants that differ between $h$ and $h_c$ can be removed due to the definition of the entropy up to a constant.\\

\noindent{\bf Acknowledgment}\\
This  research was supported by the Croatian Science
Foundation Project No. IP-2020-02-9614 \textit{Search for Quantum spacetime in Black Hole QNM spectrum and Gamma Ray Bursts}. TJ thanks H.Nikoli\'c for reading the manuscript.

\end{document}